# Mechanical equilibrium of two-component gas mixtures

Gorbunov A.A.[1], Kryukov A.P.[2], Levashov V.Yu.[2a)]

[1]*Institute for Problems in Mechanics of the Russian Academy of Sciences, prospect Vernadskogo 101, block 1, Moscow, 119526, Russia*

[2]*Department of Low Temperatures, Moscow Power Engineering Institute, Krasnokazarmennaya 14, Moscow, 111250, Russia*

Evaporation of liquid into a vapor-gas mixture is interesting and important process as for fundamental research and for different applications. In general, the rate at which a liquid evaporates is dependent on the transport of mass and energy occurring within both the liquid and vapor phases. This study is carried out on the base of continuum conservation equations for two-component medium. At the beginning the stability conditions for mechanical equilibrium are deduced. This paper is devoted to this theme.

## I. INTRODUCTION

It is known that at the presence of temperature non-uniformity in liquid or gas the convection flows can appear [1, 2]. These flows transfer mass, momentum and energy. Recent researches have shown that in some conditions the formation of stable structures: rollers, cells, circulations counters and others are possible [3]. These phenomena take place due to thermo-gravity convections. The flows and the processes occurred owing to concentration convection have investigated much less. This type of convection can be realized in gases mixtures due to the components densities difference in studied volume. Such a situation can happen at water evaporation in wetted air, for example.

At the study of water evaporation into vapor-gas (air) mixture we have found that simple estimation of the value of mass flux density made in accordance with Fick diffusion law in one dimensional approximation by two times less then the results of the experiments. Hence, there is another vapor motion which gives additional mass transfer from this liquid vapor-gas mixture interface. Our attempts to find solution of this problem in corresponding researches (see, for example, [4, 5]) had no success.

---

[a)] Corresponding author: Krasnokazarmennaya 14, Moscow 111250, Russia; Tel: +7 (495) 362-78-41; Fax.: +7 (495) 362-72-15; *E-mail address*: LevashovVY@yandex.ru

Let us note that mechanical equilibrium stability condition without taking into account viscosity influence for arbitrary state equation was presented in [1]. The equilibrium stability condition for non-compressible liquid was expressed in [3]. Equilibrium stability condition for perfect gas taking into account viscosity influence on the base of partial estimations [7] was given in paper [6]. Different correlations of mechanical equilibrium stability condition for gas at arbitrary state equation taking into account viscosity influence have been suggested in paper [8]. Nevertheless in all above mentioned papers thermo-gravity convection was considered but not convection owing to concentration difference.

The main aim of present paper is to obtain the solutions of continuum conservation equations for two-component medium in mechanical equilibrium.

## II. PROBLEM STATEMENT

Two-component mixture (for example gas and vapor) is placed in rectangle parallelepiped $\Pi$. It is assumed that in right Cartesian coordinate system $x, y, z$ segment $[0, L_x]$, $[0, L_y]$, $[0, L_z]$ determine correspondingly length $L_x$, width $L_y$, and height $L_z$ of this parallelepiped.

## III. MATHEMATICAL DESCRIPTION

### A. Main equations

Based on [9], let us consider that mixture evolution is determined by the following equations system

$$\rho\left[\rho\left(\frac{\partial V}{\partial t} + (V,\nabla)V\right) + \nabla p + \rho g e_z\right] = \mu_a\left(\rho_a + \frac{\mu_b}{\mu_a}\rho_b\right)\left(\nabla^2 V + \frac{1}{3}\nabla(\nabla,V)\right), \quad (1)$$

$$\frac{\partial \rho}{\partial t} + (V,\nabla)\rho + \rho(\nabla,V) = 0, \quad (2)$$

$$\rho\left[\frac{\partial \rho_b}{\partial t} + (V,\nabla)\rho_b + \rho_b(\nabla,V)\right] = D\left(\rho\nabla^2\rho_b - \rho_b\nabla^2\rho\right), \quad (3)$$

$$C_V^a\left(\rho_a + \rho_b\frac{C_V^b}{C_V^a}\right)\left(\frac{\partial T}{\partial t} + (V,\nabla)T\right) + p(\nabla,V) = \lambda\nabla^2 T, \quad (4)$$

$$p = R_a\left(\rho_a + \frac{R_b}{R_a}\rho_b\right)T, \tag{5}$$

$$\rho = \rho_a + \rho_b. \tag{6}$$

In this system equation (1) is the Navier - Stokes momentum equation for mixture of gases (a) and (b), equation (2) is the equation of continuity, equation (3) is the diffusion equation for gas component (b), equation (4) is the equation of internal energy conservation, equation (5) is the state equation for mixture (Dalton's law), correlation (6) presents mixture density as the sum of partial (a) and (b) components densities. In this equations system (1)-(6) $\rho$ is the mixture density; $V = (V_x, V_y, V_z)$ is the projections of mixture velocity in $x, y, z$ coordinate axes; $p$ is the mixture pressure, $g$ is the gravitation acceleration, $\mu_a, \mu_b$ are the dynamical viscosities of gas (a) and (b) correspondingly, $D$ is the diffusion coefficient, $C_V^a, C_V^b$ are the heat capacities per mass unit of gas (a) and (b), $T$ is the mixture temperature, $\lambda$ is the coefficient of thermal conductivity, $R_a, R_b$ are the individual gas (a) and (b) constants.

$\nabla \equiv \frac{\partial}{\partial x}e_x + \frac{\partial}{\partial y}e_y + \frac{\partial}{\partial z}e_z$ is the Hamilton's operator, $e_x, e_y, e_z$ are the unit orts of the coordinate system.

**B. System of dimensionless equations**

The transformation of equations system (1)-(6) to dimensionless variables is made below. Further right (roman) type – dimensionless values, variables nomenclature is same as before.

$$\rho\left[M_a^2\rho\left(Sh\frac{\partial V}{\partial t} + (V, \nabla)V\right) + \frac{1}{\gamma_a}\nabla p + C_F^a\rho e_z\right] =$$
$$= M_a C_R^a\left(\rho_a + \frac{\mu_b}{\mu_a}\rho_b\right)\left(\nabla^2 V + \frac{1}{3}\nabla(\nabla, V)\right) \tag{7}$$

$$Sh\frac{\partial \rho}{\partial t} + (V, \nabla)\rho + \rho(\nabla, V) = 0, \tag{8}$$

$$M_a Sc_a \rho \left[ Sh \frac{\partial \rho_b}{\partial t} + (V, \nabla) \rho_b + \rho_b (\nabla, V) \right] = C_R^a \left( \rho \nabla^2 \rho_b - \rho_b \nabla^2 \rho \right), \quad (9)$$

$$M_a Pr_0^a \left[ \left( \rho_a + \rho_b \frac{C_V^b}{C_V^a} \right) \left( Sh \frac{\partial T}{\partial t} + (V, \nabla) T \right) + (\gamma_a - 1) p (\nabla, V) \right] = C_R^a \nabla^2 T, \quad (10)$$

$$p = \left( \rho_a + \frac{R_b}{R_a} \rho_b \right) T, \quad (11)$$

$$\rho = \rho_a + \rho_b \quad (12)$$

Here: $V = (V_x, V_y, V_z)$, $\tilde{p} = R_a \tilde{\rho} \tilde{T}$ - pressure scale, $\tilde{\rho}$ - density scale, $\tilde{T}$ - temperature scale, $M_a = \tilde{V} / \sqrt{\gamma_a R_a \tilde{T}}$ - Mach number for gas (a), $\tilde{V}$ - velocity scale, $\gamma_a = (R_a + C_V^a) / C_V^a$ - adiabatic parameter for gas (a), $Sh = L / \tilde{V} \tilde{t}$ - Struhal number, $L$ - length scale, $\tilde{t}$ - time scale, $C_F^a = M_a^2 / Fr = gL / \gamma_a R_a \tilde{T}$, $Fr = \tilde{V}^2 / gL$ - Froude number for mixture, $C_R^a = M_a / Re_a = \mu_a / \tilde{\rho} L \sqrt{\gamma_a R_a \tilde{T}}$, $Re_a = \tilde{\rho} \tilde{V} L / \mu_a$ - Reynolds number for gas (a), $Sc_a = \mu_a / \tilde{\rho} D$ - Schmidt number for gas(a), $Pr_0^a = \mu_a C_V^a / \lambda$ - gas (a) Prandtl number.

**IV. MECHANICAL EQUILIBRIUM**

Following to [8], let us determine the mechanical equilibrium of equations system (7)-(12). It is assumed that mixture of gases (a) and (b) does not move, thus everywhere in parallelepiped $\Pi$ velocity is equal zero:

$$V = 0. \quad (13)$$

If this velocity value is substituted in dimensionless equations (7)-(12) (see above), then the following equations set is obtained

$$\frac{1}{\gamma_a} \nabla p + C_F^a \rho e_z = 0, \quad (14)$$

$$\frac{\partial \rho}{\partial t} = 0, \quad (15)$$

$$M_a \rho Sh \frac{\partial \rho_b}{\partial t} = D_a \left( \rho \nabla^2 \rho_b - \rho_b \nabla^2 \rho \right), \qquad (16)$$

$$M_a Pr_0^a \left( \rho_a + \rho_b \frac{C_V^b}{C_V^a} \right) Sh \frac{\partial T}{\partial t} = C_R^a \nabla^2 T, \qquad (17)$$

$$p = \left( \rho_a + \frac{R_b}{R_a} \rho_b \right) T, \qquad (18)$$

$$\rho = \rho_a + \rho_b. \qquad (19)$$

where in (16)

$$D_a = \frac{C_R^a}{Sc_a}.$$

It is seen from equation (15) that mixture density can depend on only space coordinates. Supposing that density of component (b) depend only on space coordinates also

$$\frac{\partial \rho_b}{\partial t} = 0, \qquad (20)$$

from (15), (19), (20) we shall have

$$\frac{\partial \rho_a}{\partial t} = 0. \qquad (21)$$

Further for simplicity let us believe that everywhere in parallelepiped $\Pi$ temperature is constant

$$T = T_m = const. \qquad (22)$$

Assumption (22) satisfies equation (17).

It is followed from (15), (18), (20)-(22) that pressure can depend only on space coordinates.

It is followed from equation (14):

$$\frac{\partial p}{\partial x} = 0, \qquad (23)$$

$$\frac{\partial p}{\partial y} = 0, \qquad (24)$$

$$\frac{1}{\gamma_a} \frac{\partial p}{\partial z} + C_F^a \rho = 0, \qquad (25)$$

It is followed from equations subsystem (23), (24) pressure depends only on coordinate $z$, that is

$$p = p(z). \tag{26}$$

It is followed from equations subsystem (23), (25)

$$\frac{\partial}{\partial z} \frac{1}{\gamma_a} \frac{\partial p}{\partial x} = 0,$$

$$\frac{\partial}{\partial x} \frac{1}{\gamma_a} \frac{\partial p}{\partial z} + C_F^a \frac{\partial}{\partial x} \rho = 0,$$

thus consistency conditions of subsystem (23), (25) has view

$$\frac{\partial \rho}{\partial x} = 0. \tag{27}$$

It is followed from subsystem (24), (25)

$$\frac{\partial}{\partial z} \frac{1}{\gamma_a} \frac{\partial p}{\partial y} = 0,$$

$$\frac{\partial}{\partial y} \frac{1}{\gamma_a} \frac{\partial p}{\partial z} + C_F^a \frac{\partial}{\partial y} \rho = 0,$$

thus consistency conditions of subsystem (24), (25) has view

$$\frac{\partial \rho}{\partial y} = 0. \tag{28}$$

It is followed from system (27), (28), that mixture density can depend only on coordinate $z$, hence:

$$\rho = \rho(z). \tag{29}$$

From (18), (19), (26), (27) it is followed

$$\frac{\partial \rho_a}{\partial x} + \frac{R_b}{R_a} \frac{\partial \rho_b}{\partial x} = 0, \tag{30}$$

$$\frac{\partial \rho_a}{\partial x} + \frac{\partial \rho_b}{\partial x} = 0, \tag{31}$$

hence, in accordance with (30), (31)

$$\frac{\partial \rho_a}{\partial x} = 0, \quad \frac{\partial \rho_b}{\partial x} = 0. \tag{32}$$

It is followed from (18), (19), (26), (28)

$$\frac{\partial \rho_a}{\partial y} + \frac{R_b}{R_a}\frac{\partial \rho_b}{\partial y} = 0, \tag{33}$$

$$\frac{\partial \rho_a}{\partial y} + \frac{\partial \rho_b}{\partial y} = 0, \tag{34}$$

hence, in accordance with (33), (34)

$$\frac{\partial \rho_a}{\partial y} = 0, \quad \frac{\partial \rho_b}{\partial y} = 0. \tag{35}$$

It is followed from system (32), (35), that densities of mixture components can depend only on coordinate $z$,

$$\rho_a = \rho_a(z), \quad \rho_b = \rho_b(z). \tag{36}$$

On the base of (26), (29), (36) subsystem of equations (14), (16) can be written as below

$$\frac{1}{\gamma_a}\frac{dp}{dz} + C_F^a \rho = 0, \tag{37}$$

$$\rho\frac{d^2\rho_b}{dz^2} - \rho_b\frac{d^2\rho}{dz^2} = 0. \tag{38}$$

Correlation for pressure (18) is substituted in equation (37). If at this (19) is taken into account, then according (22) the following equation is valid

$$T_m\left(1 - \frac{R_b}{R_a}\right)\frac{d\rho_b}{dz} = T_m\frac{d\rho}{dz} + \gamma_a C_F^a \rho. \tag{39}$$

In equation (39) the following denotations are made

$$a = 1 - \frac{R_b}{R_a}, \quad b = \frac{\gamma_a C_F^a}{T_m}. \tag{40}$$

Taking into account denotations (40), system of equations (38), (39) is written in general form

$$\varphi_1 \equiv \rho\frac{d^2\rho_b}{dz^2} - \rho_b\frac{d^2\rho}{dz^2} = 0, \tag{41}$$

$$\varphi_2 \equiv a\frac{d\rho_b}{dz} - \frac{d\rho}{dz} - b\rho = 0. \tag{42}$$

It is known the formal method of solution of equations canonical system that is system in general form which can be solved concerning the oldest derivatives of all functions in this system. This method is based on the position that canonical system is developed to system of equations of first order solved concerning the derivatives of found functions namely to system having normal Cauchy form. The equations system in normal form can be developed to one differential equation. The order of this equation is equal to quantity of equations in this normal system.

Formally equations system (41), (42) cannot be developed to canonical form: this system cannot be solved concerning $\frac{d^2\rho_b}{dz^2}, \frac{d^2\rho}{dz^2}$, because Jacobian of corresponding transformation is equal to zero, namely.

$$\frac{D(\varphi_1, \varphi_2)}{D\left(\frac{d^2\rho_b}{dz^2}, \frac{d^2\rho}{dz^2}\right)} = \begin{vmatrix} \rho & -\rho_b \\ 0 & 0 \end{vmatrix} \equiv 0.$$

However, there are at least three methods of equations (41), (42) system transformation to equations systems of general form which can be developed to canonical form. In Appendix different possible solutions of equations set (41), (42) are deduced and obtained. Content of this appendix is presented below.

In item A1 the equations (42) are solved in respect to $\frac{d\rho_b}{dz}$ and 7 different solutions of corresponding canonic system have been obtained.

In item A2 the equations (42) are solved in respect to $\frac{d\rho}{dz}$ and 8 different solutions of corresponding canonic system have been obtained. Seven from these eight solutions coincide with solutions of item A1.

In item A3 special view of equation (41) is used and 8 different solutions of corresponding system have been found in normal form. Five from these eight solutions coincide with solutions of item A1. One from these eight solutions coincides with solution of item A2.

In item A4 the solution of system (41),(42) has been created .This solution satisfies all conditions of solvability obtained in previous items.

In item A5 classification of all solutions obtained in previous items is suggested and selection of nontrivial solutions has been made.

Particularly in Appendix special trivial general solution has been obtained. This solution satisfies to condition (a36) and corresponds to mechanical equilibrium of one-component "mixture", namely:

$$\rho = C_1 e^{-bz}, \quad \rho_b = 0, \tag{43}$$

where $C_1$ - arbitrary constant. For determination $C_1$ the correlation (44) below is used

$$\int_0^{L_z} \rho dz = \rho_m L_z, \tag{44}$$

where $\rho_m$ - mean density. From (43), (44) we have

$$\rho = \frac{b\rho_m L_z}{1 - e^{bL_z}} e^{-bz}. \tag{45}$$

At

$$b \ll 1, \tag{46}$$

from (45) it is followed, that

$$\rho \approx \rho_m (1 - bz), \quad \rho_b = 0. \tag{47}$$

In accordance with Appendix general nontrivial solution of system (41),(42) takes place

$$\rho = D_1 Sh D_0 z + D_2 Ch D_0 z, \tag{48}$$
$$\rho_b = D_3 Sh D_0 z + D_4 Ch D_0 z, \tag{49}$$
$$D_3 = \frac{1}{aD_0}(D_0 D_1 + bD_2), \quad D_4 = \frac{1}{aD_0}(D_0 D_2 + bD_1). \tag{50}$$

This solution depends on three constants $D_0, D_1, D_2$, which satisfy conditions of solvability (a7), (a19), (a39), (a65), (a71), (a90), (a99).

For determination of constants in (48)-(50) solution we will use besides correlation (44) the following relationship

$$\int_0^{L_z} \rho_b dz = \rho_b^m L_z, \tag{51}$$

where $\rho_b^m$ - mean density of gas component (b).

In accordance with (44), (51) for solution of (48)-(50) we have

$$D_1 = -\frac{D_0 L_z}{2b(Ch D_0 L_z - 1)}\left(b\rho_m(Ch D_0 L_z - 1) + D_0(\rho_m - a\rho_b^m) Sh D_0 L_z\right), \tag{52}$$

$$D_2 = \frac{D_0 L_z}{2b(ChD_0 L_z - 1)} \left( D_0 (\rho_m - a\rho_b^m)(ChD_0 L_z - 1) + b\rho_m ShD_0 L_z \right), \quad (53)$$

$$D_3 = -\frac{L_z}{2ab(ChD_0 L_z - 1)} \times$$

$$\times \left( abD_0 \rho_b^m (ChD_0 L_z - 1) + \left( D_0^2 (\rho_m - a\rho_b^m) - b^2 \rho_m \right) ShD_0 L_z \right), \quad (54)$$

$$D_4 = \frac{L_z}{2ab(ChD_0 L_z - 1)} \times$$

$$\times \left( \left( D_0^2 (\rho_m - a\rho_b^m) - b^2 \rho_m \right)(ChD_0 L_z - 1) + abD_0 \rho_b^m ShD_0 L_z \right). \quad (55)$$

For determination $D_0$ in (52)-(55) let us consider(believe)

$$\frac{d\rho_b}{dz}\bigg|_{z=L_z} = 0. \quad (56)$$

From (54)-(56) the equation below is followed

$$Th\frac{D_0 L_z}{2} = \frac{D_0^2 (\rho_m - a\rho_b^m) - b^2 \rho_m}{abD_0 \rho_b^m}. \quad (57)$$

For condition (46) in accordance with (57), (52)-(55) we have (see also Appendix) the estimations

$$\rho \approx -b\rho_m \frac{\rho_m - a\rho_b^m}{\rho_m - a\rho_b^m \left(1 + \frac{bL_z}{2}\right)} z + \rho_m, \quad (58)$$

$$\rho_b \approx -\frac{b^2 \rho_m \rho_b^m L_z}{2\left(\rho_m - a\rho_b^m \left(1 + \frac{bL_z}{2}\right)\right)} z + \rho_b^m. \quad (59)$$

Supposing in (58), (59) $\rho_b^m = 0$, we obtain the estimation (47). In this case $\rho_m = \rho_a^m$, where $\rho_a^m$ means density of gas component (a).

## V. CONCLUSION

It is shown that for compressible two-component mixture of gases ((a) and (b)) at constant temperature in mechanical equilibrium different solutions of conservation equations system, namely the Navier - Stokes momentum equation for mixture, the equation of continuity and the diffusion equation are possible. First of them is trivial solution. In accordance with this solution partial density of one component, for example (b), is equal zero in whole investigated domain, but density of mixture depends on coordinate alone gravitational acceleration. In general case this dependence is exponential function of coordinate. For partial statement at small enough size of investigated volume or low gravitation this function is transformed in linear dependence.

Besides trivial case general nontrivial solution is obtained in the form of hyperbolic functions (see expressions (48) and (49) in the presented text). These solutions depend on three constants. At small size of investigated volume or low gravitation hyperbolic functions are transformed in linear dependences (58) and (59) for mixture density and partial density of one component correspondingly. In asymptotic case at zero partial density of one component this solution is transformed in the first trivial type.

**ACKNOWLEDGMENTS**

This work was supported by Russian Foundation for Basic Research (Project No. 11-08-00724)

**APPENDIX**

**A1. Equation (42) solvability concerning $\dfrac{d\rho_b}{dz}$**

Equation (42) solvability concerning $\dfrac{d\rho_b}{dz}$ depends on partial derivative

$$\frac{\partial \varphi_2}{\partial \left(\dfrac{d\rho_b}{dz}\right)} = a \,. \tag{a1}$$

**A1.1. Special solution**

Let us believe that $a = 0$ in correlation (a1). Hence equation (42) cannot be solved concerning $\dfrac{d\rho_b}{dz}$. In this case we have from (42)

$$\frac{d\rho}{dz} + b\rho = 0,$$

thus

$$\rho = C_1 e^{-bz}, \tag{a2}$$

where $C_1$ - arbitrary constant. It is followed from (41), (a2)

$$\frac{d^2\rho_b}{dz^2} - b^2\rho_b = 0,$$

hence

$$\rho_b = D_1 e^{bz} + D_2 e^{-bz}, \tag{a3}$$

where $D_1, D_2$ - arbitrary constants.

**A1.2. System of general form**

If in (a1) $a \neq 0$, then (a4) (see below) is followed from equation (41)

$$\varphi_1 = (\rho - a\rho_b)\frac{d^2\rho}{dz^2} + b\rho\frac{d\rho}{dz} = 0. \tag{a4}$$

Thus system of general form (a4), (42) takes place. System (a4), (42) solvability concerning $\frac{d^2\rho}{dz^2}, \frac{d\rho_b}{dz}$ depends on Jacobian

$$\frac{D(\varphi_1, \varphi_2)}{D\left(\frac{d^2\rho}{dz^2}, \frac{d\rho_b}{dz}\right)} = \begin{vmatrix} \rho - a\rho_b & 0 \\ 0 & a \end{vmatrix} = a(\rho - a\rho_b). \tag{a5}$$

**A1.2.1. Special solution**

Let us believe that in (a5) $\rho - a\rho_b = 0$, therefore system (a4), (42) cannot be solved concerning $\frac{d^2\rho}{dz^2}, \frac{d\rho_b}{dz}$. Then from (42) we have solution

$$\rho = 0, \quad \rho_b = 0. \tag{a6}$$

**A1.2.2. Canonical system and system in normal form**

If in (a5)
$$a(\rho - a\rho_b) \neq 0, \tag{a7}$$

then system (a4), (42) is solved concerning $\frac{d^2\rho}{dz^2}, \frac{d\rho_b}{dz}$, and we have system in canonical view

$$\frac{d^2\rho}{dz^2} = b\frac{\rho\frac{d\rho}{dz}}{a\rho_b - \rho}, \tag{a8}$$

$$\frac{d\rho_b}{dz} = \frac{1}{a}\frac{d\rho}{dz} + \frac{1}{a}b\rho. \tag{a9}$$

Denoting $r = \frac{d\rho}{dz}$, system in normal form is obtained from (a8), (a9)

$$\frac{d\rho}{dz} = r, \tag{a10}$$

$$\frac{dr}{dz} = b\frac{\rho r}{a\rho_b - \rho}, \tag{a11}$$

$$\frac{d\rho_b}{dz} = \frac{1}{a}r + \frac{1}{a}b\rho. \tag{a12}$$

### A1.2.2.1. Solvability concerning $\rho_b, r$

z differentiation (a10) and using (a11) give the possibility to analyze the following system

$$f_1 \equiv \frac{d\rho}{dz} - r = 0, \tag{a13}$$

$$f_2 \equiv \frac{d^2\rho}{dz^2} - b\frac{\rho r}{a\rho_b - \rho} = 0. \tag{a14}$$

System (a13), (a14) solvability concerning $\rho_b, r$ depends on Jacobian

$$\frac{D(f_1, f_2)}{D(\rho_b, r)} = \begin{vmatrix} 0 & -1 \\ \dfrac{ab\rho r}{(a\rho_b - \rho)^2} & -\dfrac{b\rho}{a\rho_b - \rho} \end{vmatrix} = \frac{ab\rho r}{(a\rho_b - \rho)^2}. \tag{a15}$$

#### A1.2.2.1.1. Special solutions
Let in (a15)
$$\rho r = 0. \tag{a16}$$
Hence system (a13), (a14) cannot be solved concerning $\rho_b, r$.

#### A1.2.2.1.1.1.
Let us believe that in (a16) $\rho = 0$, then from (a10) $r = 0$. Therefore from (a12) the solution is valid
$$\rho = 0, \quad \rho_b = D_1, \tag{a17}$$
where $D_1$ is arbitrary constant.

#### A1.2.2.1.1.2.
Let in (a16) at $\rho \neq 0, r = 0$. Then from (a10), (a12) solution (a18) is obtained
$$\rho = C_1, \quad \rho_b = \frac{b}{a}C_1 z + D_1, \tag{a18}$$
where $C_1, D_1$ are arbitrary constants.

#### A1.2.2.1.2. Main solution
If in (a15)
$$\rho r = \rho \frac{d\rho}{dz} \neq 0, \tag{a19}$$
then system (a13), (a14) is solved concerning $\rho_b, r$, and result is following

$$\rho_b = \frac{\rho}{a} \frac{\dfrac{d^2\rho}{dz^2} + b\dfrac{d\rho}{dz}}{\dfrac{d^2\rho}{dz^2}}. \tag{a20}$$

z differentiation (a14) and using (a10), (a20), give

$$\rho \frac{d^3\rho}{d\rho^3} - \frac{d\rho}{dz}\frac{d^2\rho}{dz^2} = \rho^2 \frac{\rho \frac{d^3\rho}{d\rho^3} - \frac{d\rho}{dz}\frac{d^2\rho}{dz^2}}{\rho^2} = \rho^2 \frac{d}{dz}\left(\frac{\frac{d^2\rho}{dz^2}}{\rho}\right) = 0,$$

thus

$$\frac{d^2\rho}{dz^2} = \alpha\rho, \qquad (a21)$$

where $\alpha \neq 0$ is arbitrary constant. From (a20), (a21) we have

$$\rho = C_1 e^{\sqrt{\alpha}z} + C_2 e^{-\sqrt{\alpha}z}, \qquad (a22)$$

$$\rho_b = \frac{1}{a}\left(C_1\left(1 + \frac{b}{\sqrt{\alpha}}\right)e^{\sqrt{\alpha}z} + C_2\left(1 - \frac{b}{\sqrt{\alpha}}\right)e^{-\sqrt{\alpha}z}\right), \qquad (a23)$$

where $C_1, C_2$ are arbitrary constants.

### A1.2.2.2. Solvability concerning $\rho, \rho_b$

z differentiation (a11) and using (a10), (a12) give the possibility to analyze the system

$$f_1 \equiv \frac{dr}{dz} - b\frac{\rho r}{a\rho_b - \rho} = 0. \qquad (a24)$$

$$f_2 \equiv \frac{d^2r}{dz^2} - b\frac{r^2}{a\rho_b - \rho} = 0, \qquad (a25)$$

System (a24), (a25) solvability concerning $\rho, \rho_b$ depends on Jacobian

$$\frac{D(f_1, f_2)}{D(\rho, \rho_b)} = \begin{vmatrix} \frac{ab\rho_b}{(a\rho_b - \rho)^2} & -\frac{ab\rho r}{(a\rho_b - \rho)^2} \\ \frac{br^2}{(a\rho_b - \rho)^2} & -\frac{abr^2}{(a\rho_b - \rho)^2} \end{vmatrix} = -\frac{ab^2 r^3}{(a\rho_b - \rho)^3}. \qquad (a26)$$

### A1.2.2.2.1. Special solution

Let in (a26) $r = 0$, thus system (a24), (a25) cannot be solved concerning $\rho, \rho_b$, and then the solution (a18) is valid.

### A1.2.2.2.2. Main solution

If in (a26) $r \neq 0$, hence system (a24), (a25) is solved concerning $\rho, \rho_b$, and then

$$\rho = r\frac{\frac{dr}{dz}}{\frac{d^2r}{dz^2}}, \qquad (a27)$$

$$\rho_b = \frac{1}{a}\frac{r}{\frac{d^2r}{dz^2}}\left(\frac{dr}{dz} + br\right), \tag{a28}$$

z differentiation (a24) and using (a10)-(a12), (a27), (a28) lead to

$$\frac{d^3r}{dz^3} = \frac{d^2r}{dz^2}\frac{\frac{dr}{dz}}{r},$$

thus

$$\frac{d^2r}{dz^2} = \alpha r, \tag{a29}$$

where $\alpha \neq 0$ is arbitrary constant. From (a29) we have

$$r = C_1 e^{\sqrt{\alpha}z} + C_2 e^{-\sqrt{\alpha}z}, \tag{a30}$$

where $C_1, C_2$ are arbitrary constants. From (a30), (a27), (a28) we have

$$\rho = \frac{1}{\sqrt{\alpha}}\left(C_1 e^{\sqrt{\alpha}z} - C_2 e^{-\sqrt{\alpha}z}\right), \tag{a31}$$

$$\rho_b = \frac{1}{a\sqrt{\alpha}}\left(C_1\left(1 + \frac{b}{\sqrt{\alpha}}\right)e^{\sqrt{\alpha}z} - C_2\left(1 - \frac{b}{\sqrt{\alpha}}\right)e^{-\sqrt{\alpha}z}\right). \tag{a32}$$

Solution (a31), (a32) corresponds to solution (a22), (a23).

### A1.2.2.3. Solvability concerning $r, \rho$

z differentiation (a12) and using (a10), (a11) give the possibility to consider the system

$$f_1 \equiv \frac{d\rho_b}{dz} - \frac{1}{a}r - \frac{b}{a}\rho = 0. \tag{a33}$$

$$f_2 \equiv \frac{d^2\rho_b}{dz^2} - br\frac{\rho_b}{a\rho_b - \rho} = 0, \tag{a34}$$

Solvability (a33), (a34) concerning $r, \rho$ depends on Jacobian

$$\frac{D(f_1, f_2)}{D(r, \rho)} = \begin{vmatrix} -\frac{1}{a} & -\frac{b}{a} \\ -b\frac{\rho_b}{a\rho_b - \rho} & -br\frac{\rho_b}{a\rho_b - \rho} \end{vmatrix} = -\frac{b}{a}\rho_b\frac{b(a\rho_b - \rho) - r}{(a\rho_b - \rho)^2}. \tag{a35}$$

### A1.2.2.3.1. Special solutions

Let in (a35)

$$\rho_b(b(a\rho_b - \rho) - r) = 0, \tag{a36}$$

thus system (a33), (a34) cannot be solved concerning $r, \rho$.

### A1.2.2.3.1.1.

At $\rho_b = 0$, in (a36) solution of system (a10)-(a12) is correlations (43).

**A1.2.2.3.1.2.**

At $\rho_b \neq 0$, $b(a\rho_b - \rho) - r = 0$, in (a36) from (a11) is followed

$$\rho = C_1 e^{bz} + C_2 e^{-bz},\tag{a37}$$

where $C_1, C_2$ are arbitrary constants. In accordance with (a12)

$$\rho_b = \frac{2}{a} C_1 e^{bz}.\tag{a38}$$

**A1.2.2.3.2. Main solution**

If in (a35)

$$\rho_b \left( b(a\rho_b - \rho) - \frac{d\rho}{dz} \right) \neq 0,\tag{a39}$$

and system (a33), (a34) is solved concerning $r, \rho$, then

$$r = a \frac{d\rho_b}{dz} - b\rho,\tag{a40}$$

$$\rho = a\rho_b \frac{\dfrac{d^2\rho_b}{dz^2} - b\dfrac{d\rho_b}{dz}}{\dfrac{d^2\rho_b}{dz^2} - b^2 \rho_b}.\tag{a41}$$

z differentiation (a34) and using (a10)-(a12), (a40), (a41) give

$$\frac{d^2\rho_b}{dz^2} = \alpha \rho_b,\ \alpha \neq b^2,\ \alpha \neq 0,\tag{a42}$$

where $\alpha$ is arbitrary constant. From (a42) we have

$$\rho_b = D_1 e^{\sqrt{\alpha} z} + D_2 e^{-\sqrt{\alpha} z},\tag{a43}$$

where $D_1, D_2$ are arbitrary constants. From (a41) one can obtain

$$\rho = \frac{a\alpha}{\alpha - b^2} \left( D_1 \left(1 - \frac{b}{\sqrt{\alpha}}\right) e^{\sqrt{\alpha} z} + D_2 \left(1 + \frac{b}{\sqrt{\alpha}}\right) e^{-\sqrt{\alpha} z} \right).\tag{a44}$$

Solution (a43), (a44) corresponds solution (a22), (a23).

**A2. System of general type**

The equation (42) is solved concerning $\dfrac{d\rho}{dz}$, because $\dfrac{\partial \varphi_2}{\partial \left(\dfrac{d\rho}{dz}\right)} = -1$. Thus (a45) is followed from equation (41)

$$\varphi_1 = (\rho - a\rho_b) \frac{d^2\rho_b}{dz^2} + b\rho_b \frac{d\rho}{dz} = 0,\tag{a45}$$

Hence system of general type (a45), (42) is deduced. System (a45), (42) solvability concerning $\dfrac{d^2\rho_b}{dz^2}, \dfrac{d\rho}{dz}$ depends on Jacobian

$$\dfrac{D(\varphi_1, \varphi_2)}{D\left(\dfrac{d^2\rho_b}{dz^2}, \dfrac{d\rho}{dz}\right)} = \begin{vmatrix} \rho - a\rho_b & 0 \\ 0 & -1 \end{vmatrix} = a\rho_b - \rho. \qquad (a46)$$

### A2.1. Special solutions

Let in (a46)
$$a\rho_b - \rho = 0, \qquad (a47)$$
thus system (a45), (42) cannot be solved concerning $\dfrac{d^2\rho_b}{dz^2}, \dfrac{d\rho}{dz}$.

### A2.1.1.

At $a = 0$, in (a46) in accordance with (a45), (42) we have solution
$$\rho = 0, \quad \rho_b = \psi(z), \qquad (a48)$$
where $\psi(z)$ is arbitrary function of coordinate $z$.

### A2.1.2.

At $a \neq 0$ in (a46) one can obtain $\rho = a\rho_b$ and from (42) the solution (a6) is valid.

### A2.2. Canonical system and system in normal form

If in (a46) $a\rho_b - \rho \neq 0$, then system (a45), (42) is solved concerning $\dfrac{d^2\rho_b}{dz^2}, \dfrac{d\rho}{dz}$, and as a result the following system in canonical view is deduced

$$\dfrac{d^2\rho_b}{dz^2} = -b\dfrac{\rho_b \dfrac{d\rho}{dz}}{\rho - a\rho_b}, \qquad (a49)$$

$$\dfrac{d\rho}{dz} = a\dfrac{d\rho_b}{dz} - b\rho. \qquad (a50)$$

Introducing denotation, $r = \dfrac{d\rho_b}{dz}$, from (a49), (a50) we obtain system in normal form

$$\dfrac{d\rho_b}{dz} = r, \qquad (a51)$$

$$\dfrac{d\rho}{dz} = ar - b\rho. \qquad (a52)$$

$$\dfrac{dr}{dz} = -b\dfrac{\rho_b}{\rho - a\rho_b}(ar - b\rho), \qquad (a53)$$

### A2.2.1 Solvability concerning $r, \rho$

z differentiation (a51) and using (a53)) give the possibility to analyze the system

$$f_1 \equiv \frac{d\rho_b}{dz} - r = 0, \tag{a54}$$

$$f_2 \equiv \frac{d^2\rho_b}{dz^2} + b\frac{\rho_b}{\rho - a\rho_b}(ar - b\rho) = 0. \tag{a55}$$

Solvability of system (a54), (a55) concerning $r, \rho$ depends on Jacobian.

$$\frac{D(f_1, f_2)}{D(r, \rho)} = \begin{vmatrix} -1 & 0 \\ \dfrac{ab\rho_b}{\rho - a\rho_b} & -\dfrac{ab\rho_b(r - b\rho_b)}{(\rho - a\rho_b)^2} \end{vmatrix} = \frac{ab\rho_b(r - b\rho_b)}{(\rho - a\rho_b)^2}. \tag{a56}$$

**A2.2.1.1. Special solutions**
Let in (a56)

$$a\rho_b(r - b\rho_b) = 0, \tag{a57}$$

thus system (a54), (a55) cannot be solved concerning $r, \rho$.

**A2.2.1.1.1.**
At $a = 0$, in (a57) from (a52), (a53) solution (a2), (a3) is valid.

**A2.2.1.1.2.**
At $a \neq 0, \rho_b = 0$, in (a57) from (a52) solution (43) is valid.

**A2.2.1.1.3.**
At $a \neq 0, \rho_b \neq 0, r - b\rho_b = 0$, in (a57) we have

$$\rho_b = D_1 e^{bz}, \tag{a58}$$

where $D_1$ is arbitrary constant. From (a50)

$$\rho = \frac{a}{2} D_1 e^{bz} + C_1 e^{-bz}, \tag{a59}$$

where $C_1$ is arbitrary constant.
Solution (a58), (a59) corresponds to solution (a37), (a38).

**A2.2.1.2. Main solution**
If in (a56)

$$a\rho_b \left( \frac{d\rho_b}{dz} - b\rho_b \right) \neq 0, \tag{a60}$$

then system (a54), (a55) is solved concerning $r, \rho$, and equality (a41) takes place. z differentiation (a55) and using (a51)-(a53), (a41) give the possibility to obtain the equation (a42) and, consequently, solution (a43), (a44).

**A2.2.2. Solvability concerning $r, \rho_b$**
z differentiation (a52) and using (a53) give the possibility to consider system

$$f_1 \equiv \frac{d\rho}{dz} - ar + b\rho = 0, \tag{a61}$$

$$f_2 \equiv \frac{d^2\rho}{dz^2} + b\frac{\rho}{\rho - a\rho_b}(ar - b\rho) = 0. \tag{a62}$$

Solvability of system (a61), (a62) concerning $r, \rho_b$ depends on Jacobian

$$\frac{D(f_1, f_2)}{D(r, \rho_b)} = \begin{vmatrix} -a & 0 \\ \frac{ab\rho}{\rho - a\rho_b} & \frac{ab\rho(ar - b\rho)}{(\rho - a\rho_b)^2} \end{vmatrix} = -\frac{a^2 b\rho(ar - b\rho)}{(\rho - a\rho_b)^2}. \tag{a63}$$

**A2.2.2.1. Special solutions**

Let in (a63)

$$a\rho(ar - b\rho) = 0, \tag{a64}$$

thus system (a61), (a62) cannot be solved concerning $r, \rho_b$.

**A2.2.2.1.1.**

At $a = 0$, in (a64) solution (a2), (a3) is valid.

**A2.2.2.1.2.**

At $a \neq 0, \rho = 0$, in (a64) from (a53) solution (a17) is valid.

**A2.2.2.1.3.**

At $a \neq 0, \rho \neq 0, ar - b\rho = 0$, in (a64) from (a53) solution (a18) is valid.

**A2.2.2.2. Main solution**

If in (a63)

$$a\rho\left(a\frac{d\rho_b}{dz} - b\rho\right) \neq 0, \tag{a65}$$

then system (a61), (a62) is solved concerning $r, \rho_b$, and from (a61) one can obtain

$$r = \frac{1}{a}\left(\frac{d\rho}{dz} + b\rho\right), \tag{a66}$$

from (a62) correlation (a20) is followed. z differentiation (a62) and using (a51)-(a53), (a66), (a20) give the equation (a21) and solution (a22), (a23).

**A2.2.3. Solvability concerning $\rho, \rho_b$**

z differentiation (a53) and using (a51)-(a53) give the possibility to analyze system

$$f_1 \equiv \frac{dr}{dz} + \frac{b\rho_b}{\rho - a\rho_b}(ar - b\rho) = 0, \tag{a67}$$

$$f_2 \equiv \frac{d^2 r}{dz^2} + b\frac{r}{\rho - a\rho_b}(ar - b\rho) = 0. \tag{a68}$$

Solvability of system (a67), (a68) concerning $\rho, \rho_b$ depends on Jacobian

$$\frac{D(f_1, f_2)}{D(\rho, \rho_b)} = \begin{vmatrix} ab\rho_b \dfrac{b\rho_b - r}{(\rho - a\rho_b)^2} & b\rho \dfrac{ar - b\rho}{(\rho - a\rho_b)^2} \\ abr \dfrac{b\rho_b - r}{(\rho - a\rho_b)^2} & abr \dfrac{ar - b\rho}{(\rho - a\rho_b)^2} \end{vmatrix} =$$

$$= -\frac{ab^2 r(b\rho_b - r)(ar - b\rho)}{(\rho - a\rho_b)^3}. \tag{a69}$$

**A2.2.3.1. Special solutions**

Let in (a69)

$$ar(b\rho_b - r)(ar - b\rho) = 0, \tag{a70}$$

thus system (a67), (a68) cannot be solved concerning $\rho, \rho_b$

**A2.2.3.1.1.**

At $a = 0$, in (a70) solution (a2), (a3) is valid.

**A2.2.3.1.2.**

At $a \neq 0, r = 0$, in (a70) from (a51), (a53) solution (a17) is valid.

**A2.2.3.1.3.**

At $a \neq 0, r \neq 0, b\rho_b - r = 0$, in (a70) solution (a58), (a59) is valid.

**A2.2.3.1.4.**

At $a \neq 0, r \neq 0, ar - b\rho = 0$, in (a70) from (a53) solution (a18) is valid.

**A2.2.3.2. Main solution**

If in (a69)

$$a\frac{d\rho_b}{dz}\left(b\rho_b - \frac{d\rho_b}{dz}\right)\left(a\frac{d\rho_b}{dz} - b\rho\right) \neq 0, \tag{a71}$$

then system (a67), (a68) is solved concerning $\rho, \rho_b$, and

$$\rho_b = r \frac{\dfrac{dr}{dz}}{\dfrac{d^2 r}{dz^2}}, \tag{a72}$$

$$\rho = ar \frac{\dfrac{dr}{dz} - br}{\dfrac{d^2 r}{dz^2} - b^2 r}. \tag{a73}$$

z differentiation (a68) and using (a51)-(a53), (a72), (a73) give the equation (a29), and, consequently, solution

$$\rho_b = \frac{1}{\sqrt{\alpha}}\left(D_1 e^{\sqrt{\alpha}z} - D_2 e^{-\sqrt{\alpha}z}\right), \tag{a74}$$

$$\rho = a\left(\frac{1}{\sqrt{\alpha}+b}D_1 e^{\sqrt{\alpha}z} - \frac{1}{\sqrt{\alpha}-b}D_2 e^{-\sqrt{\alpha}z}\right). \tag{a75}$$

Solution corresponds solution (a22), (a23).

### A3. Special view of equation (41)

From equation (41) we have

$$0 = \rho\frac{d^2\rho_b}{dz^2} - \rho_b\frac{d^2\rho}{dz^2} = \rho\frac{d^2\rho_b}{dz^2} + \frac{d\rho_b}{dz}\frac{d\rho}{dz} - \rho_b\frac{d^2\rho}{dz^2} - \frac{d\rho}{dz}\frac{d\rho_b}{dz} =$$

$$= \frac{d}{dz}\left(\rho\frac{d\rho_b}{dz} - \rho_b\frac{d\rho}{dz}\right). \tag{a76}$$

(a77) is followed from (a76)

$$\varphi_1 \equiv \rho\frac{d\rho_b}{dz} - \rho_b\frac{d\rho}{dz} - C_0 = 0, \tag{a77}$$

where $C_0$ is arbitrary constant, thus system of general form (a77), (42) takes place.

Solvability of system (a77), (42) concerning $\dfrac{d\rho_b}{dz}, \dfrac{d\rho}{dz}$ depends on Jacobian.

$$\frac{D(\varphi_1,\varphi_2)}{D\left(\dfrac{d\rho_b}{dz}, \dfrac{d\rho}{dz}\right)} = \begin{vmatrix} \rho & -\rho_b \\ a & -1 \end{vmatrix} = a\rho_b - \rho. \tag{a78}$$

### A3.1. Special solutions

Let in (a78)

$$a\rho_b - \rho = 0, \tag{a79}$$

thus system (a77), (42) cannot be solved concerning $\dfrac{d\rho_b}{dz}, \dfrac{d\rho}{dz}$.

#### A3.1.1.

At $a = 0$, in (a79) from (a77) $C_0 = 0$, and, consequently, solution (a48) is valid.

#### A3.1.2.

At $a \neq 0$, in (a79) $\rho = a\rho_b$, $C_0 = 0$, and from (42) solution (a6) is valid.

### A3.2. System in normal form

If in (a78) $a\rho_b - \rho \neq 0$, then system (a77), (42) is solved concerning $\dfrac{d\rho_b}{dz}, \dfrac{d\rho}{dz}$, and from (a77), (42) we have system in normal form.

$$f_1 \equiv \frac{d\rho}{dz} - \frac{b\rho^2 - aC_0}{a\rho_b - \rho} = 0, \tag{a80}$$

$$f_2 \equiv \frac{d\rho_b}{dz} - \frac{b\rho\rho_b - C_0}{a\rho_b - \rho} = 0. \tag{a81}$$

### A3.2.1. Solvability concerning $\rho_b$

Solvability of equation (a80) concerning $\rho_b$ depends on partial derivative

$$\frac{\partial f_1}{\partial \rho_b} = -\frac{a(b\rho^2 - aC_0)}{(a\rho_b - \rho)^2}. \tag{a82}$$

#### A3.2.1.1. Special solutions

Let in (a82)

$$a(b\rho^2 - aC_0) = 0, \tag{a83}$$

thus equation (a80)) cannot be solved concerning $\rho_b$.

#### A3.2.1.1.1.

At $a = 0$, in (a83) from (a80), (a81) one can deduce

$$\rho = C_1 e^{-bz}, \quad \rho_b = e^{-bz}\left(\frac{C_0}{C_1}\frac{1}{2b}e^{2bz} + C_2\right). \tag{a84}$$

Solution (a84) corresponds to solution (a2), (a3).

#### A3.2.1.1.2.

Let in (a83) $a \neq 0$, $b\rho^2 - aC_0 = 0$, hence

$$\rho^2 = \frac{aC_0}{b} > 0. \tag{a85}$$

#### A3.2.1.1.2.1.

If in (a85)

$$\rho = \sqrt{\frac{aC_0}{b}}, \tag{a86}$$

Then from (a81)

$$\rho_b = \frac{b}{a}\sqrt{\frac{aC_0}{b}}z + C_1, \tag{a87}$$

where $C_1$ is arbitrary constant. Solution (a86), (a87) corresponds to solution (a18).

#### A3.2.1.1.2.2.

If in (a84)

$$\rho = -\sqrt{\frac{aC_0}{b}}, \tag{a88}$$

Then from (a81)

$$\rho_b = -\frac{b}{a}\sqrt{\frac{aC_0}{b}}z + C_1, \tag{a89}$$

where $C_1$ is arbitrary constant.

### A3.2.1.2. Main solution
If in (a82)
$$a(b\rho^2 - aC_0) \neq 0, \tag{a90}$$
then equation (a80) is solved concerning $\rho_b$, and
$$\rho_b = \frac{\rho}{a} + \frac{b\rho^2 - aC_0}{a\dfrac{d\rho}{dz}}. \tag{a91}$$
z differentiation (a80) with taking into account (a81), (a91) give the equation
$$(b\rho^2 - aC_0)\frac{d^2\rho}{dz^2} - b\rho\left(\frac{d\rho}{dz}\right)^2 = 0. \tag{a92}$$
Solution of equation (a92) is found in the form
$$\rho = C_1 e^{\sqrt{\alpha}z} + C_2 e^{-\sqrt{\alpha}z}, \tag{a93}$$
where $C_1, C_2, \alpha \neq 0$ are arbitrary constants. Substitution (a93) in (a92) leads to correlation
$$\left(b\left(C_1^2 e^{2\sqrt{\alpha}z} + C_2^2 e^{-2\sqrt{\alpha}z} + 2C_1 C_2\right) - aC_0\right)\left(C_1 e^{\sqrt{\alpha}z} + C_2 e^{-\sqrt{\alpha}z}\right)\alpha -$$
$$- b\left(C_1 e^{\sqrt{\alpha}z} + C_2 e^{-\sqrt{\alpha}z}\right)\left(C_1^2 e^{2\sqrt{\alpha}z} + C_2^2 e^{-2\sqrt{\alpha}z} - 2C_1 C_2\right)\alpha = 0,$$
Hence
$$4bC_1 C_2 - aC_0 = 0,$$
or
$$\rho = C_1 e^{\sqrt{\alpha}z} + \frac{1}{C_1}\frac{aC_0}{4b} e^{-\sqrt{\alpha}z}. \tag{a94}$$
From (a91), (a94) we have
$$\rho_b = \frac{1}{a}\left(C_1\left(1 + \frac{b}{\sqrt{\alpha}}\right)e^{\sqrt{\alpha}z} + \frac{1}{C_1}\frac{aC_0}{4b}\left(1 - \frac{b}{\sqrt{\alpha}}\right)e^{-\sqrt{\alpha}z}\right). \tag{a95}$$
Solution (a94), (a95) corresponds to solution (a22), (a23).

### A3.2.2. Solvability concerning $\rho$
Solvability of equation (a81) concerning $\rho$ depends on partial derivative
$$\frac{\partial f_2}{\partial \rho} = -\frac{ab\rho_b^2 - C_0}{(a\rho_b - \rho)^2}. \tag{a96}$$

### A3.2.2.1. Special solutions
Let in (a96)
$$ab\rho_b^2 - C_0 = 0, \tag{a97}$$
thus equation (a81) cannot be solved concerning $\rho$.

### A3.2.2.1.1.

At $a = 0$, in (a97) from (a80), (a81) the following solution is valid

$$\rho = C_1 e^{-bz}, \quad \rho_b = D_1 e^{-bz}, \qquad (a98)$$

where $C_1, D_1$ are arbitrary constants.

**A3.2.2.1.2.**

At $a \neq 0, \rho_b = 0$, in (a97) the solution (43) is valid.

**A3.2.2.2. Main solution**

If in (a96)

$$ab\rho_b^2 - C_0 \neq 0, \qquad (a99)$$

then equation (a81) is solved concerning $\rho$, and

$$\rho = \frac{a\rho_b \dfrac{d\rho_b}{dz} + C_0}{\dfrac{d\rho_b}{dz} + b\rho_b}. \qquad (a100)$$

z differentiation (a81) with taking into account (a80), (a100) give the equation

$$(ab\rho_b^2 - C_0)\frac{d^2\rho_b}{dz^2} - ab\rho_b\left(\frac{d\rho_b}{dz}\right)^2 + b^2 C_0 \rho_b = 0. \qquad (a101)$$

Solution of equation (a101) is found in the form

$$\rho_b = C_1 e^{\sqrt{\alpha} z} + C_2 e^{-\sqrt{\alpha} z}, \qquad (a102)$$

where $C_1, C_2, \alpha \neq 0$ are arbitrary constants. Substitution (a102) in (a101) leads to correlation

$$\left(ab\left(C_1^2 e^{2\sqrt{\alpha} z} + C_2^2 e^{-2\sqrt{\alpha} z} + 2C_1 C_2\right) - C_0\right)\left(C_1 e^{\sqrt{\alpha} z} + C_2 e^{-\sqrt{\alpha} z}\right)\alpha -$$
$$- ab\left(C_1 e^{\sqrt{\alpha} z} + C_2 e^{-\sqrt{\alpha} z}\right)\left(C_1^2 e^{2\sqrt{\alpha} z} + C_2^2 e^{-2\sqrt{\alpha} z} - 2C_1 C_2\right)\alpha +$$
$$+ b^2 C_0 \left(C_1 e^{\sqrt{\alpha} z} + C_2 e^{-\sqrt{\alpha} z}\right) = 0.$$

Hence

$$4ab\alpha C_1 C_2 = C_0\left(\alpha - b^2\right),$$

and

$$\rho_b = C_1 e^{\sqrt{\alpha} z} + \frac{1}{C_1}\frac{C_0\left(\alpha - b^2\right)}{4ab\alpha} e^{-\sqrt{\alpha} z}. \qquad (a103)$$

In accordance with (a100) taking into account (a103)

$$\rho = \frac{\sqrt{\alpha} a}{\sqrt{\alpha} + b} C_1 e^{\sqrt{\alpha} z} + \frac{1}{C_1}\frac{C_0\left(\sqrt{\alpha} + b\right)}{4\sqrt{\alpha} b} e^{-\sqrt{\alpha} z}. \qquad (a104)$$

Solution (a103), (a104) corresponds to solution (a22), (a23).

**A4. Solution of system (41),(42), satisfying all conditions of solvability**

Taking into account conditions of solvability (a7), (a19), (a39), (a60), (a65), (a71), (a90), (a99) let us find solutions of equations system (41), (42) in the form

$$\rho = C_1 e^{\sqrt{\alpha}z} + C_2 e^{-\sqrt{\alpha}z}, \quad \rho_b = C_3 e^{\sqrt{\alpha}z} + C_4 e^{-\sqrt{\alpha}z}, \tag{a105}$$
where $C_1, C_2, C_3, C_4, \alpha \neq 0$ are arbitrary constants.

### A4.1. Hyperbolic solution

Let $\alpha > 0, D_0 = \sqrt{\alpha}$. In accordance with (a105) we have (48), (49). Substitution (48), (49) in equation (41)) leads to correlation
$$(D_1 ShD_0 z + D_2 ChD_0 z)D_0^2(D_3 ShD_0 z + D_4 ChD_0 z) -$$
$$- (D_3 ShD_0 z + D_4 ChD_0 z)D_0^2(D_1 ShD_0 z + D_2 ChD_0 z) = 0,$$
wich is satisfied with identify. Substitution (48), (49) in equation (42) gives
$$aD_0(D_3 ChD_0 z + D_4 ShD_0 z) = (D_0 D_1 + bD_2)ChD_0 z + (D_0 D_2 + bD_1)ShD_0 z,$$
from this correlation (50) is followed, thus
$$\rho_b = \frac{1}{aD_0}((D_0 D_1 + bD_2)ShD_0 z + (D_0 D_2 + bD_1)ChD_0 z). \tag{a106}$$
In this case solution (a22), (a23) corresponds to solution (48), (a106).

### A4.2. Trigonometric solution

Let $\alpha < 0, \sqrt{\alpha} = iD_0$. Let us believe in accordance with (a105)
$$\rho = D_1 SinD_0 z + D_2 CosD_0 z, \tag{a107}$$
$$\rho_b = D_3 SinD_0 z + D_4 CosD_0 z, \tag{a108}$$
where $D_1, D_2, D_3, D_4$ are arbitrary constants. Substitution (a107), (a108) in equation (41) leads to correlation
$$-(D_1 SinD_0 z + D_2 CosD_0 z)D_0^2(D_3 SinD_0 z + D_4 CosD_0 z) +$$
$$+ (D_3 SinD_0 z + D_4 CosD_0 z)D_0^2(D_1 SinD_0 z + D_2 CosD_0 z) = 0,$$
which is satisfied with identify. Substitution (a107), (a108) in equation (42) gives
$$aD_0(D_3 CosD_0 z - D_4 SinD_0 z) =$$
$$= (D_0 D_1 + bD_2)CosD_0 z + (D_0 D_2 - bD_1)SinD_0 z,$$
hence
$$D_3 = \frac{1}{aD_0}(D_0 D_1 + bD_2), \quad D_4 = \frac{1}{aD_0}(D_0 D_2 - bD_1),$$
and
$$\rho_b = \frac{1}{aD_0}((D_0 D_1 + bD_2)SinD_0 z + (D_0 D_2 - bD_1)CosD_0 z). \tag{a109}$$
In this case solution (a22), (a23) corresponds to solution (a107), (a109).

### A5. Solutions classification and some particular solutions

Let us give the name "trivial" solutions of equations (41), (42), which satisfy the following correlation
$$\rho \rho_b \equiv 0. \tag{a110}$$

Correspondingly solutions, which do not satisfy to (a110) we will name as "nontrivial."
Let us consider solutions corresponding to the following conditions

$$\rho > 0, \tag{a111}$$

$$\rho_b > 0, \tag{a112}$$

$$\rho - \rho_b > 0, \tag{a113}$$

**A5.1. Solutions not satisfying conditions to (a111)-(a113)**

Solutions, which do not satisfy to condition (a111), are the following trivial solutions: (a6); (a17); (a48).

Nontrivial solution (a88), (a89) does not satisfy to condition (a111).

Solution, which does not satisfy to condition (a112), is trivial solution (43).

**A5.2. Particular solutions**

For determination of constants values in nontrivial solutions correlations (44), (51) are used.

**A5.2.1. Solutions depending on two constants**

Solutions, depending on two arbitrary constants, are (a18); (a37), (a38); (a98).

**A5.2.1.1. Solution (a18)**

In accordance with (44), (51) for solution (a18) we have

$$C_1 = \rho_m, \quad D_1 = \rho_b^m - \frac{b}{a}\rho_m L_z,$$

or

$$\rho = \rho_m, \tag{a114}$$

$$\rho_b = \frac{b}{a}\rho_m\left(z - \frac{L_z}{2}\right) + \rho_b^m, \tag{a115}$$

$$\rho - \rho_b = -\frac{b}{a}\rho_m\left(z - \frac{L_z}{2}\right) + \rho_m - \rho_b^m. \tag{a116}$$

In accordance with (a114) condition (a111) takes place. From (a112), (a113), (a115), (a116) it is followed

$$\frac{b}{a}\rho_m\left(z - \frac{L_z}{2}\right) + \rho_b^m > 0, \quad -\frac{b}{a}\rho_m\left(z - \frac{L_z}{2}\right) + \rho_m - \rho_b^m > 0. \tag{a117}$$

At $z = 0$, from (a117) the expression (a118) is deduced

$$-\frac{b}{a}\rho_m\frac{L_z}{2} + \rho_b^m > 0, \quad \frac{b}{a}\rho_m\frac{L_z}{2} + \rho_m - \rho_b^m > 0. \tag{a118}$$

At $z = L_z$, from (a117) the following correlations are valid

$$\frac{b}{a}\rho_m\frac{L_z}{2} + \rho_b^m > 0, \quad -\frac{b}{a}\rho_m\frac{L_z}{2} + \rho_m - \rho_b^m > 0. \tag{a119}$$

From (118), (a119) it is followed that (a112), (a113) are satisfied, if $|a| > bL_z$.

**A5.2.1.2. Solution (a37), (a38)**

In accordance with (44), (51) for solution (a37), (a38) we have

$$C_1 = \frac{a}{2} \frac{\rho_b^m L_z}{e^{bL_z} - 1}, \quad C_2 = \frac{\rho_m - \frac{a}{2}\rho_b^m}{e^{-bL_z} - 1} L_z,$$

and

$$\rho = \frac{a}{2} \frac{\rho_b^m L_z}{e^{bL_z} - 1} e^{bz} + \frac{\rho_m - \frac{a}{2}\rho_b^m}{e^{-bL_z} - 1} L_z e^{-bz}, \tag{a120}$$

$$\rho_b = \frac{\rho_b^m L_z}{e^{bL_z} - 1} e^{bz}, \tag{a121}$$

$$\rho - \rho_b = \left(\frac{a}{2} - 1\right) \frac{\rho_b^m L_z}{e^{bL_z} - 1} e^{bz} + \frac{\rho_m - \frac{a}{2}\rho_b^m}{e^{-bL_z} - 1} L_z e^{-bz}, \tag{a122}$$

From (a121) it is followed that, condition (a112) takes place. From (a111)), (a113) (a120), (a122) the following correlations can be obtained

$$\frac{a}{2} \frac{\rho_b^m}{e^{bL_z} - 1} e^{bz} + \frac{\rho_m - \frac{a}{2}\rho_b^m}{e^{-bL_z} - 1} e^{-bz} > 0, \tag{a123}$$

$$\left(\frac{a}{2} - 1\right) \frac{\rho_b^m}{e^{bL_z} - 1} e^{bz} + \frac{\rho_m - \frac{a}{2}\rho_b^m}{e^{-bL_z} - 1} e^{-bz} > 0, \tag{a124}$$

From (a123)

$$a > \frac{2\rho_m e^{bL_z}}{\rho_b^m \left(1 + e^{bL_z}\right)}. \tag{a125}$$

From (a124)

$$a > \frac{2\left(\rho_m e^{bL_z} + \rho_b^m\right)}{\rho_b^m \left(1 + e^{bL_z}\right)}. \tag{a126}$$

From (a125), (a126) it is followed that (a111)), (a113) are satisfied if (a126) is valid.

### A5.2.1.3. Solution (a98)

In accordance (44), (51) for solution (a98) we have

$$C_1 = \rho_m \frac{bL_z}{1 - e^{-bL_z}}, \quad D_1 = \rho_b^m \frac{bL_z}{1 - e^{-bL_z}},$$

and correspondingly

$$\rho = \rho_m \frac{bL_z}{1 - e^{bL_z}} e^{-bz}, \tag{a127}$$

$$\rho_b = \rho_b^m \frac{bL_z}{1 - e^{bL_z}} e^{-bz}, \tag{a128}$$

$$\rho - \rho_b = \left(\rho_m - \rho_b^m\right) \frac{bL_z}{1 - e^{bL_z}} e^{-bz}, \tag{a129}$$

From (a127)-(a129) it is followed, that conditions (a111)-(a113) take place, because $\rho_m - \rho_b^m = \rho_a^m > 0$, where $\rho_a^m$ - mean gas (a) density.

### A5.2.2. Solutions depending on three constants
Solutions depending on three arbitrary constants are (a2), (a3); (48), (a106).

### A5.2.2.1. Solution (a2), (a3)
In accordance with (44), (50) for solution (a2), (a3) correlations (a130), (a130) are valid

$$C_1 = \frac{b\rho_m L_z}{1 - e^{-bL_z}}, \qquad (a130)$$

$$D_1 = \frac{b\rho_b^m L_z}{e^{bL_z} - 1} - e^{-bL_z} D_2. \qquad (a131)$$

Let us believe that condition (56) takes place. From (56) it is followed that

$$D_1 = \frac{e^{-bL_z}}{e^{bL_z}} D_2. \qquad (a132)$$

From (a131), (a132)

$$D_1 = \frac{b\rho_b^m L_z e^{-bL_z}}{2 ShbL_z}, \quad D_2 = \frac{b\rho_b^m L_z e^{bL_z}}{2 ShbL_z}. \qquad (a133)$$

In accordance with (a130), (a133)

$$\rho = \frac{b\rho_m L_z}{1 - e^{-bL_z}} e^{-bz}, \qquad (a134)$$

$$\rho_b = \frac{b\rho_b^m L_z}{ShbL_z} Chb(L_z - z). \qquad (a135)$$

From (a134), (a135) it is followed, that (a111), (a112) are satisfied. Also from (a134), (a135) at $b \ll 1$, the estimations are valid

$$\rho \approx \rho_m(1 - bz), \quad \rho_b \approx \rho_b^m,$$

andr

$$\rho - \rho_b \approx (\rho_m - \rho_b^m) - b\rho_m z. \qquad (a136)$$

Estimation (a136) at $b \ll 1$, provides satisfaction (a113).

### A5.2.2.2. Solution (48), (a106)
In accordance with (44), (50) for solution (48), (a106), taking into account (50), we have correlations (52)-(55). Let correlation (56) is satisfied. From (56), (50), (a106), (54), (55) it is followed

$$-\left(abD_0\rho_b^m(ChD_0L_z - 1) + \left(D_0^2(\rho_m - a\rho_b^m) - b^2\rho_m\right)ShD_0L_z\right)ChD_0L_z +$$
$$+ \left(\left(D_0^2(\rho_m - a\rho_b^m) - b^2\rho_m\right)(ChD_0L_z - 1) + abD_0\rho_b^m ShD_0L_z\right)ShD_0L_z = 0,$$

and (57) takes place. Equation (57) has solution if

$$\left|\frac{D_0^2(\rho_m - a\rho_b^m) - b^2\rho_m}{abD_0\rho_b^m}\right| < 1. \qquad (a137)$$

From (a137) at

$$0 < a < \frac{\rho_m}{\rho_b^m}, D_0 > 0, \quad (a138)$$

non-equality (a139) takes place

$$b < D_0 < b\frac{\rho_m}{\rho_m - a\rho_b^m}, \quad (a139)$$

and at

$$a < 0, D_0 < 0, \quad (a140)$$

$$-b\frac{\rho_m}{\rho_m - a\rho_b^m} < D_0 < -b. \quad (a141)$$

For estimation (46) from (57) taking into account (a138)-(a141)

$$D_0 = \pm\beta, \quad \beta = b\sqrt{\frac{\rho_m}{\rho_m - a\rho_b^m\left(1 + \frac{bL_z}{2}\right)}}, \quad (a142)$$

From (46), (52)-(55) one can obtain

$$D_1 \approx -\frac{D_0}{b}(\rho_m - a\rho_b^m), \quad D_2 \approx \rho_m, \quad (a143)$$

$$D_3 \approx -\frac{1}{aD_0}\left(\frac{D_0^2}{b}(\rho_m - a\rho_b^m) - b\rho_m\right), \quad D_4 \approx \rho_b^m, \quad (a144)$$

From (46), (a142)-(a144), (48), (a106) we have (58), (59), and also

$$\rho - \rho_b \approx -b\rho_m \frac{\rho_m - a\rho_b^m}{\rho_m - a\rho_b^m\left(1 + \frac{bL_z}{2}\right)} z + \rho_m - \rho_b^m. \quad (a145)$$

For (58) condition (a111) is satisfied, if

$$b < \frac{\rho_m - a\rho_b^m}{\left(\rho_m - \frac{a\rho_b^m}{2}\right)L_z}.$$

For (59) condition (a112) is satisfied, if

$$b < \frac{\sqrt{(a\rho_b^m)^2 + 8\rho_m(\rho_m - a\rho_b^m)} - a\rho_b^m}{2\rho_m L_z}.$$

For (a145) condition (a113) is satisfied, if

$$b < \frac{(\rho_m - a\rho_b^m)(\rho_m - \rho_b^m)}{\left(\rho_m(\rho_m - a\rho_b^m) + \frac{a\rho_b^m}{2}(\rho_m - \rho_b^m)\right)L_z}.$$

**REFERENCES**

[1] Landau L.D., Lifschitz E.M. *Fluid Mechanics* (Pergamon, London, 1959).

[2] Isachenko V.P., Osipova V.A., Sukomel A.S. *Heat Transfer*. Moscow. Energia. 1969 (In Russian).

[3] Gershuni G.Z., Guhovitcki E.M. *Convective stability of non-compressible liquid*. Moscow. Nauka. 1972 (In Russian).

[4] G.D. McBain, H. Suehrcke, J.A. Harris, "Evaporation from an open cylinder," International Journal of Heat and Mass Transfer **43** 2117 (2000).

[5] P. L. Kelly-Zion, C. J. Pursell, R. S. Booth, A. N. VanTilburg, "Evaporation rates of pure hydrocarbon liquids under the influence of natural convection and diffusion," International Journal of Heat and Mass Transfer **52** 3305 (2009).

[6] Polezhaev V.I., Vlasuk M.P. "About cell convection in infinity long horizontal gas layer heated below," Dokladi AN SSSR **195** 1058 (1970) (in Russian).

[7] Jeffreys H. "The stability of a compressible fluid heated below." Proc. Cambr. Phil. Soc. 26. 1930. P. 170-172.

[8] Gorbunov A.A. About the stability of gas mechanical equilibrium in confined volumes, Izvestiya vuzov, North-Caucasian region, Natural Sciences, Issue "Actual problems of mathematical hydrodynamics" 73 (2009). (in Russian).

[9] Labuntsov D.A., Yagov V.V., Kryukov A.P. *Bases of two-phase systems mechanics*. Moscow. Moscow Power Engineering Institute. 1988. (in Russian).